\documentclass[aps,prb,superscriptaddress,twocolumn,showpacs,preprintnumbers,amsmath,amssymb]{revtex4}
\usepackage[dvips]{graphicx,color}% Include figure files
\usepackage{dcolumn}% Align table columns on decimal point
\usepackage{bm}% bold math
\usepackage{ulem}
\usepackage{soul}
\begin{document}

\title{Relation between spin Hall effect and anomalous Hall effect \\ 
in 3$d$ ferromagnetic metals}

\author{Yasutomo~Omori}
\thanks{These authors contributed equally to this work.}
\affiliation{Institute for Solid State Physics, University of Tokyo, Kashiwa, Chiba 277-8581, Japan}
\author{Edurne~Sagasta}
\thanks{These authors contributed equally to this work.}
\affiliation{CIC nanoGUNE, 20018 Donostia-San Sebastian, Basque Country, Spain}
\author{Yasuhiro~Niimi}
\email{niimi@phys.sci.osaka-u.ac.jp}
\affiliation{Institute for Solid State Physics, University of Tokyo, Kashiwa, Chiba 277-8581, Japan}
\affiliation{Department of Physics, Graduate School of Science, Osaka University, Toyonaka, Osaka 560-0043, Japan}
\author{Martin~Gradhand}
\affiliation{H. H. Wills Physics Laboratory, University of Bristol, Bristol BS8 1TL, United Kingdom}
\author{Luis~E.~Hueso}
\affiliation{CIC nanoGUNE, 20018 Donostia-San Sebastian, Basque Country, Spain}
\affiliation{IKERBASQUE, Basque Foundation for Science, 48011 Bilbao, Basque Country, Spain}
\author{F\`elix~Casanova}
\affiliation{CIC nanoGUNE, 20018 Donostia-San Sebastian, Basque Country, Spain}
\affiliation{IKERBASQUE, Basque Foundation for Science, 48011 Bilbao, Basque Country, Spain}
\author{YoshiChika Otani}
\email{yotani@issp.u-tokyo.ac.jp}
\affiliation{Institute for Solid State Physics, University of Tokyo, Kashiwa, Chiba 277-8581, Japan}
\affiliation{RIKEN-CEMS, 2-1 Hirosawa, Wako, Saitama 351-0198, Japan}

\date{January 3, 2019}

\begin{abstract}
We study the mechanisms of the spin Hall effect (SHE) and 
anomalous Hall effect (AHE) in 3$d$ ferromagnetic metals 
(Fe, Co, permalloy (Ni$_{81}$Fe$_{19}$; Py), and Ni) 
by varying their resistivities and temperature. 
At low temperatures where the phonon scattering is negligible, 
the skew scattering coefficients of the SHE and AHE in Py 
are related to its spin polarization. However, this simple relation 
breaks down for Py at higher temperatures as well as for the other 
ferromagnetic metals at any temperature. We find that, in general, 
the relation between the SHE and AHE is more complex, 
with the temperature dependence of the SHE being much stronger than 
that of AHE. 
\end{abstract}
\maketitle

\section{Introduction}

The spin Hall effect (SHE) and its inverse (ISHE) enable us to 
interconvert spin and charge currents in the transverse direction 
and are widely recognized as essential methods to generate and 
detect spin currents in spintronic 
devices~\cite{hoffmann_review_2013,sinova_review_2015,niimi_review_2015}. 
Since the original predictions of the 
SHE~\cite{dp_phys_lett_1971,hirsch_prl_1999}, it has been 
experimentally investigated in a variety of nonmagnetic materials 
with strong spin-orbit interactions such as III-V 
semiconductors~\cite{kato_science_2004,wunderlich_prl_2005}, 
4$d$ and 5$d$ transition metals~\cite{saitoh_apl_2006,ando_prl_2008,mosendz_prl_2010,mosendz_prb_2010,liu_prl_2011,morota_prb_2011,kondou_apex_2012,isasa_prb_2015,sagasta_prb_2016}, 
alloys~\cite{gu_prl_2010,niimi_prl_2011,niimi_prl_2012,laczkowski_apl_2014}, 
oxides~\cite{fujiwara_nat_commun_2013}, and 
organic materials~\cite{ando_nat_mater_2013}. 
The mechanism of the SHE can be extrinsic or intrinsic. 
The former depends on 
the combination of the host metal and 
impurities~\cite{gradhand_prl_2010,ebert_prl_2011,fert_prl_2011}, 
while the latter depends on the detailed properties of the momentum-space 
Berry phase~\cite{guo_prl_2008,tanaka_prb_2008}. 
These mechanisms are the same as for the anomalous Hall effect (AHE) 
in ferromagnetic metals (FMs), which has been intensively studied 
for many years~\cite{nagaosa_review_2010}. Thus, it has been commonly accepted 
that the SHE shares the same origin as the AHE~\cite{martin_prb_2014}.

\begin{figure}
\begin{center}
\includegraphics[width=8.5cm]{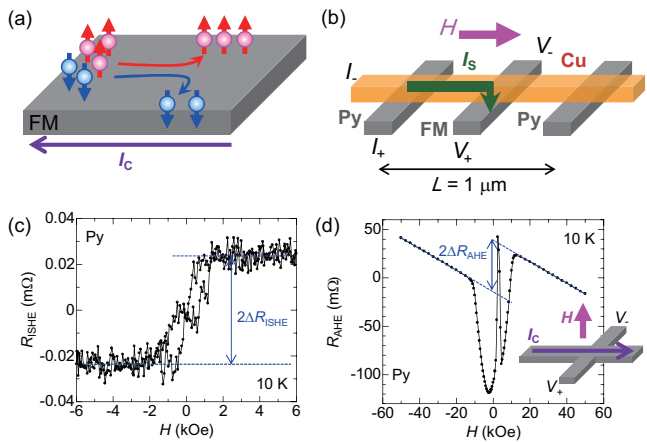}
\caption{(a) Intuitive schematic of the SHE and the AHE in FM. Spin and charge accumulations appear in the transverse direction respect to the incident current $I_{\rm C}$ and are detected as the SHE and the AHE, respectively. (b) Schematic of a lateral spin valve for the spin absorption method to measure the ISHE in FM. The magnetic field $H$ is applied along the Cu wire. (c) Inverse spin Hall resistance $R_{\rm ISHE}$ of the middle Py wire (20~nm in thickness) as a function of $H$ at 10~K using the configuration shown in (b). The ISHE signal ($\Delta R_{\rm ISHE}$) is defined in the figure. (d) Anomalous Hall resistance $R_{\rm AHE}$ of Py as a function of $H$ at 10 K. The AHE signal ($\Delta R_{\rm AHE}$) is defined in the figure. The inset shows a part of a Hall bar for the AHE measurement. The magnetic field $H$ is applied perpendicular to the plane. Compared to the SHE configuration, the field direction is rotated by 90$^{\circ}$.} \label{fig1}
\end{center}
\end{figure}

It was experimentally verified that not only 
the ISHE~\cite{miao_prl_2013,tsukahara_prb_2014,du_prb_2014} 
but also the SHE~\cite{das_prb_2017,das_nano_lett_2018,ralph_pra_2018,iihama_nature_electronics_2018,qin_prb_2017} occur in FMs with finite spin polarization. In FMs, both spin and charge accumulations 
can exist~\cite{taniguchi_pra_2015} and are detected as the SHE and the AHE, 
respectively [see Fig. 1(a)]. Very recently, the control of the spin accumulation by manipulating 
the magnetization of the ferromagnet has been achieved 
experimentally~\cite{das_prb_2017,das_nano_lett_2018,ralph_pra_2018,iihama_nature_electronics_2018} 
after its theoretical predictiont~\cite{taniguchi_pra_2015}. Das \textit{et al}.~\cite{das_prb_2017} used 
the term ``anomalous SHE" to describe this mechanism. The anomalous SHE can be understood 
by generalizing the spin conductivity $\sigma_{ij}^{s}$ to $\sigma_{ij}^{s} (M)$ allowing for 
the anisotropy of $\sigma_{ij}^{s}$ as discussed by Seemann \textit{et al}.~\cite{seemann_prb_2015}. 
Here $ij$ are the spatial directions, $s$ refers to the direction of the spin and $M$ 
is the global magnetization of the ferromagnet. For the simplest contribution to this general response, 
in other words, $M \propto s$, it was experimentally verified that the SHE also occurs in 
FMs with finite spin polarization. 
Intuitively, in FMs, both spin and charge accumulations can exist and are 
detected as the SHE and the AHE, respectively [see Fig.~\ref{fig1}(a)]. 
Thus, it was suggested that the SHE and AHE in FMs are related via the spin 
polarization~\cite{tsukahara_prb_2014}. However, it has not been 
experimentally verified if this simple relation is general, and 
therefore valid for all the FMs and all the mechanisms. 
From a theoretical viewpoint, such a relation might hold 
in the limit of diffusive transport~\cite{theory} 
but is not expected to hold in general. 

In this work, we present a detailed investigation of
the relation between the SHE 
and the AHE in four different 3$d$ FMs, i.e., 
Fe, Co, permalloy (Ni$_{81}$Fe$_{19}$; Py), and Ni. 
By changing the residual resistivity of the FM at low temperatures, 
the skew scattering contribution 
(one of the extrinsic mechanisms)~\cite{smit_physica_1958} can be separated 
from other contributions. It turns out that the aforementioned relation 
between the SHE and AHE holds for the skew scattering term in Py. 
However, this simple relation is not valid for the other mechanisms in Py 
and for the other FMs. The SHE in the 3$d$ FMs has much stronger temperature 
dependence than the AHE. 
We discuss a possible scenario to explain the observed results. 

\section{Experimental Details}

To perform the SHE measurements in the 3$d$ FMs, we adopted the spin 
absorption method in the lateral spin valve 
structure~\cite{morota_prb_2011,isasa_prb_2015,sagasta_prb_2016,niimi_prl_2011,niimi_prl_2012,laczkowski_apl_2014,fujiwara_nat_commun_2013}. 
This method enables us to estimate the spin diffusion length and 
the spin Hall angle ($\theta_{\rm SHE}$) on the same device. 
The SHE devices were fabricated on SiO$_{2}$/Si 
substrates with multiple-step electron beam lithography followed 
by metal deposition and lift-off. 
We first patterned two 100-nm-wide wires and deposited Py by 30~nm 
in thickness by electron beam evaporation. The two Py wires are 
separated by a length ($L$) of 1~$\mu$m, as illustrated in Fig.~1(b). 
One of the Py wires is used as a spin current injector, 
while the other is used to estimate the spin diffusion length of our target 
wire, as detailed in Supplemental Material~\cite{supplement}.
In the second step, 
the target 3$d$ FM wire (hereafter middle wire)
with the width ($w_{\rm M}$) of 200~nm was placed just
in the middle of the two Py wires 
and a 5- to 30-nm-thick 3$d$ FM (Fe, Co, Py, or Ni) was deposited with 
electron beam evaporation. In the third step, a 100-nm-wide and 
100-nm-thick Cu strip was bridged on top of the three wires 
with a Joule heating evaporator. Before the Cu evaporation, 
an Ar-ion milling treatment was performed 
to achieve transparent interfaces. 
For the AHE measurements, 
a 20-$\mu$m-long and 3-$\mu$m-wide Hall bar was patterned 
with electron beam lithography and the FM (5 to 30 nm in thickness) was 
deposited at the same time as the SHE devices were prepared. 
We then capped all the devices with Al$_{2}$O$_{3}$ using radio frequency 
magnetron sputtering to protect them from oxidization. 
All the electric transport measurements were performed in a $^{4}$He flow 
cryostat using the lock-in technique.

\section{Results and Discussions}

\subsection{SHE and AHE in Py}

When an electric current $I_{\rm C}$ is injected from Py to the left side 
of Cu as shown in Fig.~\ref{fig1}(b), spin accumulation is created 
at the interface and diffuses in the Cu bridge. In this process, 
a pure spin current $I_{\rm S}$ flows in the Cu channel on the right side. 
Most of $I_{\rm S}$ is then absorbed 
into the middle wire and converted into charge current via the ISHE, 
which is detected as a voltage drop $V_{\rm ISHE} (=V_{+}-V_{-})$. 
The ISHE resistance $R_{\rm ISHE} = V_{\rm ISHE}/I_{\rm C}$ is measured 
by sweeping the external magnetic field $H$ along the Cu channel. 
It is saturated when the magnetization of the Py wire is fully polarized. 
The difference of $R_{\rm ISHE}$ between the positive and negative saturated 
magnetic fields is the ISHE signal, defined as $2\Delta R_{\rm ISHE}$. 
As shown in Fig.~\ref{fig1}(c), 
a positive $\Delta R_{\rm ISHE} (\sim 25~\mu \Omega)$ 
was obtained at 10~K for a 20~nm thick Py middle wire
with the longitudinal resistivity 
$\rho_{xx}$ of 22~$\mu \Omega$$\cdot$cm. 

We also confirmed 
the reciprocity in the present system for FM. 
This can be realized by exchanging the electrodes 
($V_{+} \leftrightarrow I_{+}$, $V_{-} \leftrightarrow I_{-}$) 
on the same device and measuring the direct SHE, 
as detailed in Supplemental Material~\cite{supplement}.
It is well-known that the AHE occurs in FMs as a result of 
the breaking of time reversal symmetry. However, 
the Onsager reciprocal relation holds for the SHE 
in FMs~\cite{buttiker_prb_2012} 
because the \textit{total} number of spin-up and 
spin-down electrons is always kept constant. 

By using the spin transport model proposed 
by Takahashi and Maekawa~\cite{takahashi_maekawa_prb_2003}, 
the spin Hall resistivity 
$-\rho_{xy}^{\rm SHE}$ 
can be estimated as follows: 
\begin{eqnarray}
-\rho_{xy}^{\rm SHE} = \rho_{yx}^{\rm SHE} = \theta_{\rm SHE} \rho_{xx} 
= \frac{w_{\rm M}}{x} \left( \frac{I_{\rm C}}{\bar{I}_{\rm S}} \right) \Delta R_{\rm ISHE}
\label{eq1}
\end{eqnarray}
where $x$ is the shunting factor and $\bar{I}_{\rm S}$ 
is the effective spin current absorbed into the FM middle wire. 
When $\bar{I}_{\rm S}$ is converted into charge current in 
the middle wire, a part of the charge current is shunted by 
the Cu bridge on the middle FM wire. 
The shunting factor $x$ has been 
calculated with a finite elements method using 
SpinFlow~3D~\cite{niimi_prl_2012,sagasta_prb_2016}. 
$\bar{I}_{\rm S}$ can be determined 
from nonlocal spin valve measurements 
with and without the middle wire~\cite{supplement}. 
$\bar{I}_{\rm S}$ is also related to the spin diffusion length 
of the middle FM wire. 

We next measured the AHE with a Hall bar pattern, prepared at the same time 
as the SHE device. By applying an out-of-plane magnetic field and 
flowing $I_{\rm C}$ in the longitudinal direction of the Hall bar, 
a transverse voltage drop $V_{\rm AHE} (= V_{+}-V_{-})$ is detected, 
as sketched in the inset of Fig.~\ref{fig1}(d). Figure~~\ref{fig1}(d) 
shows a typical $R_{\rm AHE} = V_{\rm AHE}/I_{\rm C}$ vs $H$ curve 
for Py at 10 K. Although there are two backgrounds, 
namely normal Hall resistance and 
planar Hall resistance in between $\pm 10$~kOe~\cite{zhang_jap_2013}, 
a clear positive AHE signal 
$\Delta R_{\rm AHE}$ can be extracted. From $\Delta R_{\rm AHE}$, 
we obtain the anomalous Hall resistivity defined as, 
\begin{eqnarray}
-\rho_{xy}^{\rm AHE} = \rho_{yx}^{\rm AHE} = \theta_{\rm AHE}\rho_{xx} 
= t \Delta R_{\rm AHE},
\label{eq2}
\end{eqnarray}
where $\theta_{\rm AHE}$ is the anomalous Hall angle and 
$t$ is the thickness of the Hall bar. 

\begin{figure}
\begin{center}
\includegraphics[width=8.5cm]{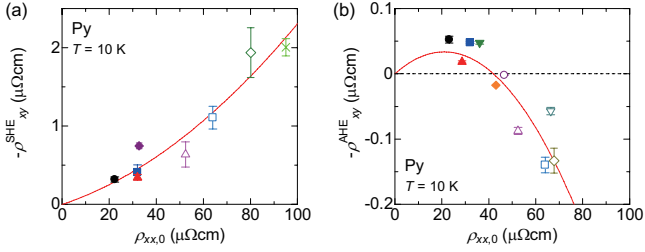}
\caption{(a) Spin Hall resistivity $-\rho_{xy}^{\rm SHE}$ and (b) anomalous Hall resistivity $-\rho_{xy}^{\rm AHE}$ in Py as a function of $\rho_{xx,0}$ at 10 K. The solid lines are the best fits with Eq. (2). The same symbol is used in (a) and (b) if the Py deposition is done at the same time for the SHE and AHE samples. The dotted line in (b) indicates $-\rho_{xy}^{\rm AHE}=0$.} \label{fig2}
\end{center}
\end{figure}

By plotting $-\rho_{xy}^{\rm SHE}$ and $-\rho_{xy}^{\rm AHE}$, obtained 
with Eqs.~(1) and (2), as a function of $\rho_{xx}$, 
the detailed mechanisms can be addressed 
as shown in previous works~\cite{sagasta_prb_2016,hou_prl_2015,tian_prl_2009}. 
For this purpose, the recent scaling equation proposed by 
Hou $\textit{et al}$.~\cite{hou_prl_2015} is useful: 
\begin{eqnarray}
-\rho_{xy}^{\rm H} = \alpha_{\rm ss}^{\rm H} \rho_{xx,0} 
&+& \beta_{0}^{\rm H}(\rho_{xx,0})^{2} \nonumber\\
&+& \gamma^{\rm H}\rho_{xx,0}\rho_{xx,T} 
+ \beta_{1}^{\rm H}(\rho_{xx,T})^{2},
\label{eq3}
\end{eqnarray}
where H refers to the SHE or AHE, $\rho_{xx,0}$ is the residual resistivity 
at low temperature (in the present case, at 10 K), 
$\rho_{xx,T} (= \rho_{xx} -\rho_{xx,0})$ is the resistivity 
induced by phonons 
and $\alpha_{\rm ss}^{\rm H}$ is the skew scattering angle 
due to impurities or grain boundaries. As detailed in 
Ref.~\onlinecite{hou_prl_2015}, the side-jump terms due to static 
(impurities or grain boundaries) and dynamic (phonons) scattering 
sources as well as the intrinsic contribution originating from the band 
structure~\cite{yang_prb_2011,hou_j_phys_2012,zhu_prb_2014,xu_sci_bull_2015,ye_prb_2012,wu_prb_2013,kotzler_prb_2005} 
are entangled in $\beta_{0}^{\rm H}$, $\gamma^{\rm H}$, 
and $\beta_{1}^{\rm H}$ in a complex manner. 
Nevertheless, as discussed in Ref.~\onlinecite{hou_prl_2015}, 
the effect of the intrinsic Berry curvature is most strongly 
reflected in the $\beta_{1}^{\rm H}$ term. 

Firstly, to simplify Eq.~(\ref{eq3}), we focus on the low temperature part 
where the phonon contribution is negligible and consider the case of Py. 
By substituting $\rho_{xx,T} = 0$ in Eq.~(\ref{eq3}), 
a simplified equation can be obtained: 
\begin{eqnarray}
-\rho_{xy}^{\rm H} = \alpha_{\rm ss}^{\rm H} \rho_{xx,0} 
+ \beta_{0}^{\rm H}(\rho_{xx,0})^{2}. 
\label{eq4}
\end{eqnarray}
In order to determine $\alpha_{\rm ss}^{\rm H}$ and 
$\beta_{0}^{\rm H}$, the SHE and AHE of Py have to be measured in a wide 
$\rho_{xx,0}$ range. For this purpose, we changed the thickness of the Py 
wire (from 5 to 30~nm) and also the deposition rate 
(from 0.04~nm/s to 0.08~nm/s), as already demonstrated in our previous work 
for Pt~\cite{sagasta_prb_2016}. Figures~\ref{fig2}(a) and \ref{fig2}(b) 
show $-\rho_{xy}^{\rm SHE}$ and $-\rho_{xy}^{\rm AHE}$ of Py at 10 K 
as a function of $\rho_{xx,0}$, respectively. $-\rho_{xy}^{\rm SHE}$ 
increases with increasing $\rho_{xx,0}$, while $-\rho_{xy}^{\rm AHE}$ 
decreases with $\rho_{xx,0}$. By fitting $-\rho_{xy}^{\rm SHE}$ and 
$-\rho_{xy}^{\rm AHE}$ with Eq.~(\ref{eq4}), the skew scattering 
term $\alpha_{\rm ss}^{\rm H}$ and the combination of the side-jump and 
intrinsic contributions $\beta_{0}^{\rm H}$ can be obtained as follows: 
$\alpha_{\rm ss}^{\rm SHE} = 1.0 \pm 0.4$\%, 
$\alpha_{\rm ss}^{\rm AHE} = 0.32 \pm 0.1$\%, 
$\beta_{0}^{\rm SHE} = 131 \pm 60$~$\Omega^{-1}\cdot$cm$^{-1}$ and 
$\beta_{0}^{\rm AHE} = -76 \pm 20$~$\Omega^{-1}\cdot$cm$^{-1}$.
$\alpha_{\rm ss}^{\rm AHE}$ and $\beta_{0}^{\rm AHE}$ are 
in good agreement with previous reports~\cite{zhang_jap_2013,note2}. 

\begin{table*}
\caption{The coefficients $\beta_{1}^{\rm H}$ and $\gamma^{\rm H}$ extracted from the fittings with Eq.~(\ref{eq3}) for each FM. For comparison, we also show the coefficient of the quadratic term of the AHE from previous works (Refs.~\onlinecite{zhang_jap_2013,hou_j_phys_2012,ye_prb_2012,wu_prb_2013,kotzler_prb_2005}) in the table.}
\label{table1}
\begin{ruledtabular}
\begin{tabular}{cccccc}
FM & $\beta_{1}^{\rm SHE}$ & $\beta_{1}^{\rm AHE}$ & $\beta_{1}^{\rm AHE}$ or $b$~\cite{note2} in literature & $\gamma^{\rm SHE}$ & $\gamma^{\rm AHE}$ \\ 
 & ($\times 10^{3}$~$\Omega^{-1}$cm$^{-1}$) & ($\times 10^{3}$~$\Omega^{-1}$cm$^{-1}$) & ($\times 10^{3}$~$\Omega^{-1}$cm$^{-1}$) & ($\times 10^{3}$~$\Omega^{-1}$cm$^{-1}$) & ($\times 10^{3}$~$\Omega^{-1}$cm$^{-1}$) \\ \hline
Fe & $4.9 \pm 0.2$ & $0.89 \pm 0.04$ & 1.1~\cite{zhang_jap_2013}, 0.82~\cite{wu_prb_2013} & $-1.1 \pm 0.1$ & $1.50 \pm 0.03$ \\
Co & $-8.3 \pm 0.5$ & $0.34 \pm 0.03$ & 0.2~\cite{kotzler_prb_2005}, 0.73~\cite{hou_j_phys_2012} & $0.04 \pm 0.24$ & $0.97 \pm 0.02$ \\
Py & $-10.1 \pm 0.3$ & $-0.056 \pm 0.015$ & $-0.05$~\cite{zhang_jap_2013} & $0.57 \pm 0.14$ & $-0.002 \pm 0.009$ \\ 
Ni & $-17.1 \pm 0.5$ & $-0.14 \pm 0.11$ & $-(0.5 \sim 1.0)$~\cite{ye_prb_2012} & $5.9 \pm 0.4$ & $-0.89 \pm 0.09$ \\
\end{tabular}
\end{ruledtabular}
\end{table*}

Interestingly, the ratio of the AHE and SHE in Py for 
the skew scattering contribution, 
$\alpha_{\rm ss}^{\rm AHE}/\alpha_{\rm ss}^{\rm SHE} = 0.32$, 
is a reasonable value for the spin polarization $p$ of 
Py~\cite{niimi_prl_2011,sagasta_apl_2017}. 
In other words, for the skew scattering, the relation between the AHE and SHE 
can be expressed as 
\begin{eqnarray}
\rho_{xy}^{\rm AHE} = p \rho_{xy}^{\rm SHE}. 
\label{eq5}
\end{eqnarray}
This can be understood intuitively as follows. In FMs, incident spin-up 
and spin-down electrons are deflected to the transverse opposite directions, 
as illustrated in Fig.~\ref{fig1}(a). Since the number of spin-up 
electrons (3 in the sketch) is larger than that of spin-down 
electrons (2 in the sketch), there is a finite charge 
accumulation (the difference of the deflected electrons, 
i.e., 1 in the sketch) along the transverse direction 
with respect to the incident current direction, which can be detected 
as an anomalous Hall voltage. On the other hand, the spin accumulation is 
proportional to the difference of spin directions (5 in the sketch), 
which can be detected as a spin Hall voltage. Thus, the ratio of 
the AHE and SHE is indeed the spin polarization 
($p=(3-2)/(3+2)=0.2$ in the sketch of Fig.~\ref{fig1}(a)).

\begin{figure}
\begin{center}
\includegraphics[width=8.5cm]{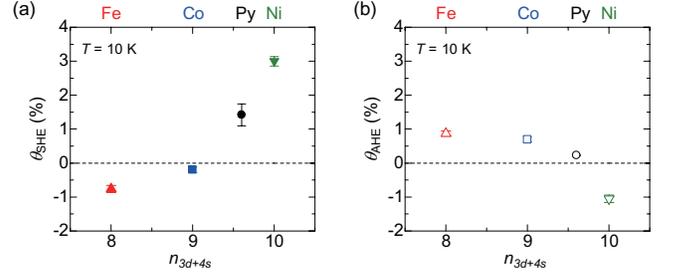}
\caption{(a) Spin Hall angle $\theta_{\rm SHE}$ and (b) anomalous Hall angle $\theta_{\rm AHE}$ measured at 10~K as a function of the number of  electrons in the outermost shell. The thickness of the the 3$d$ FMs is 20~nm. The dotted lines in (a) and (b) indicate $\theta_{\rm H}=0$.} \label{fig3}
\end{center}
\end{figure}

\begin{figure}
\begin{center}
\includegraphics[width=8.5cm]{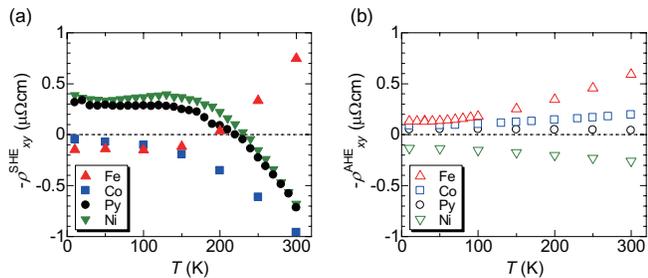}
\caption{The temperature dependence of (a) spin Hall resistivity $-\rho_{xy}^{\rm SHE}$ and (b) anomalous Hall resistivity $-\rho_{xy}^{\rm AHE}$ in four FMs (Py, Fe, Co and Ni). The thickness of the four FMs is 20 nm both for the SHE and AHE measurements.} \label{fig4}
\end{center}
\end{figure}

In fact, this simple picture can be applied for diffusive scattering systems 
such as Py. As we detail in Ref.~\onlinecite{supplement}, 
Eq.~(\ref{eq5}) can be derived in Mott's two current model~\cite{mott_1964} 
under the specific assumption 
$\displaystyle \frac{\rho_{xy}^{\uparrow}}{\rho_{xx}^{\uparrow}} = -\frac{\rho_{xy}^{\downarrow}}{\rho_{xx}^{\downarrow}}$ . Here $\rho_{ij}^{\uparrow}$ 
and $\rho_{ij}^{\downarrow}$ are the spin-up and spin-down resistivity tensor 
elements. 
Py is a random alloy composed of Ni and Fe. 
The anisotropy on the Fermi surface should be suppressed 
and lead to more isotropic scattering properties. Thus, the Hall angle 
is essentially a spin-independent property averaged over 
all the contributing states. 
This supports the finding that the simplified relation holds 
for the skew scattering in Py. 

\subsection{SHE and AHE in other 3$d$ FMs}

However, such a simple picture does not work for the other $3d$ FMs. 
The electronic states can be quite anisotropic because of 
the complicated band structure of $3d$ FMs. 
Those states should show distinct effective spin-orbit couplings 
for spin-up and spin-down electrons.
Thus, the above specific assumption can break down. 
We show $\theta_{\rm SHE}$ and $\theta_{\rm AHE}$ 
at $T=10$~K for the $3d$ FMs in Figs.~\ref{fig3}(a) and \ref{fig3}(b), 
respectively. 
As in the case of the intrinsic SHEs in $4d$ and $5d$ transition 
metals~\cite{morota_prb_2011,guo_prl_2008,tanaka_prb_2008}, 
$\theta_{\rm SHE}$ is expected to change 
the sign from negative to positive with increasing 
the number of electrons in the outer shell~\cite{du_prb_2014}. 
Such a tendency can be seen clearly in 
$\theta_{\rm SHE}$ of the $3d$ FMs in Fig.~\ref{fig3}(a).
However, the sign of $\theta_{\rm SHE}$ is 
opposite to that of $\theta_{\rm AHE}$ for Fe, Co, and Ni. 
Even in the case of Py, $\theta_{\rm AHE}$ is negative 
when $\rho_{xx,0}$ is more than 40~$\mu\Omega\cdot$cm, 
as shown in Fig.~\ref{fig2}(b).
This obviously shows that Eq.~(\ref{eq5}) is not 
general and the detailed band structure 
of the electron orbitals has to be taken into account, 
as mentioned above. 

\begin{figure*}
\begin{center}
\includegraphics[width=16cm]{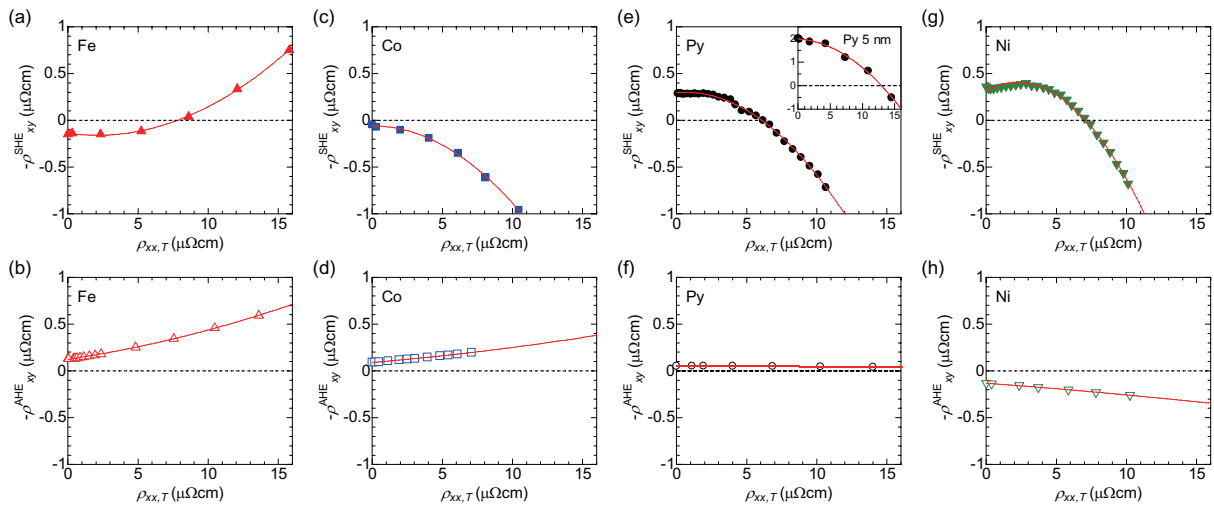}
\caption{Spin Hall resistivity $-\rho_{xy}^{\rm SHE}$ and anomalous Hall resistivity $-\rho_{xy}^{\rm AHE}$ in four FMs (Fe, Co, Py, and Ni) at finite temperatures. $-\rho_{xy}^{\rm SHE}$ as a function of $\rho_{xx,T}$ in (a) Fe, (c) Co, (e) Py, and (g) Ni. Anomalous Hall resistivity $-\rho_{xy}^{\rm AHE}$ as a function of $\rho_{xx,T}$ in (b) Fe, (d) Co, (f) Py, and (h) Ni. $\rho_{xx,T}$ varies by changing temperature from 10~K to 300~K. The solid lines are the best fits of the data to Eq.~(\ref{eq3}). The thickness of the four FMs is 20 nm both for SHE and AHE measurements except for the inset in (e). The inset in (e) shows $-\rho_{xy}^{\rm SHE}$ of 5 nm thick Py wire. For the fitting with Eq.~(\ref{eq3}), the same parameter $\beta_{1}^{\rm SHE} = -10.1 (\times 10^{3}$ $\Omega^{-1}\cdot{\rm cm}^{-1}$) was used both for 20 nm and 5 nm thick Py wires.}
\label{fig5}
\end{center}
\end{figure*}

So far, we have focused on the low temperature parts of the SHE and AHE. 
To address the effect of dynamic disorders, we next discuss 
the temperature dependences of the SHE and AHE in Figs.~\ref{fig4}(a) 
and \ref{fig4}(b), respectively. 
The temperature dependence of the SHE is much stronger than that of the AHE. 
For Fe, Py, and Ni, the sign of $-\rho_{xy}^{\rm SHE}$ is changed 
at 200-250~K, while such a sign change cannot be seen 
for $-\rho_{xy}^{\rm AHE}$.
To specify the reason for such temperature dependences, 
we have fitted both $-\rho_{xy}^{\rm SHE}$ and $-\rho_{xy}^{\rm AHE}$ 
as a function of $\rho_{xx,T}$ with Eq.~(\ref{eq3}) 
as shown in Fig.~\ref{fig5}, and obtained 
$\beta_{1}^{\rm H}$ and $\gamma^{\rm H}$ as the quadratic and linear terms 
in Eq.~(\ref{eq3}), respectively (see Table~\ref{table1}). 
For example, $\beta_{1}^{\rm SHE}$ of Py is more than two orders of 
magnitude larger than $\beta_{1}^{\rm AHE}$. Even for Py, the relations 
between the SHE and AHE for $\beta_{1}^{\rm H}$ and $\gamma^{\rm H}$ 
are not as simple as the skew scattering term.

A similar tendency can be seen for the other $3d$ FMs. 
For Fe, Co, and Ni, $\beta_{1}^{\rm SHE}$ is 
one or two orders of magnitude larger than $\beta_{1}^{\rm AHE}$, 
as shown in Table~\ref{table1}. Note that $\beta_{1}^{\rm AHE}$ values 
in the present work are in good agreement with previous experiments 
(see Table~\ref{table1}) and tight-binding calculations~\cite{naito_prb_2010}. 
On the other hand, $|\beta_{1}^{\rm SHE}|$ of the $3d$ FMs ranges 
between 4.9 and $17\times 10^{3}$~$\Omega^{-1}\cdot{\rm cm}^{-1}$, 
which is larger than that of a typical SHE material, Pt 
($1.6\times 10^{3}$~$\Omega^{-1}\cdot{\rm cm}^{-1}$)~\cite{morota_prb_2011,sagasta_prb_2016}. 
The relation between $\gamma^{\rm SHE}$ and $\gamma^{\rm AHE}$ 
strongly varies with the $3d$ FMs (see Table~\ref{table1}). 

Much larger $\beta_{1}^{\rm SHE}$ values than $\beta_{1}^{\rm AHE}$ ones 
would originate from 
the stronger temperature dependence of the SHE in 3$d$ FMs. 
At the moment, we do not have a conclusive 
picture for the origin of this dependence. 
In general, the spin transport can be mediated not only 
by conduction electrons but also 
by magnons in FMs~\cite{klaui_nat_commun_2018,van_wees_nat_phys_2015}. 
One possible scenario 
is the contribution of electron-magnon interactions in 3$d$ FMs. 
The electron-magnon interactions would induce additional spin-flip processes. 
We note that such spin-flip processes are equivalent in magnitude 
for up-to-down and down-to-up spin channels even in 
ferromagnetic systems~\cite{ref}. 
In such a situation, some asymmetric scatterings which are spin-dependent 
would contribute only to the SHE but not to the AHE, 
and thus would be associated with the fact that the strong 
temperature dependence is not present in the AHE of the $3d$ FMs or 
the SHE of nonmagnetic metals. 
Interestingly, a recent theoretical report claims that magnon spin current 
can be significant around room temperature in 3$d$ 
FMs~\cite{cheng_prb_2017}, which 
might be related to our case. 
However, there are some open questions: 
how large the asymmetric scatterings are quantitatively and 
whether any other mechanisms contribute to the observed 
spin Hall resistivity or not.
These would be addressed in future. 

\section{Summary}

In conclusion, we experimentally investigated the relation 
between the SHE and AHE in four 3$d$ 
FMs [Fe, Co, Py (Ni$_{81}$Fe$_{19}$), and Ni]. 
In a typical ferromagnetic alloy, Py, the skew scattering contribution 
of the AHE is related to that of the SHE via the spin polarization of Py, 
as can be understood intuitively. However, this relation does not hold 
for other mechanisms. This fact is highlighted by the temperature dependence 
of the SHE and AHE. For all the 3$d$ FMs, one of 
the intrinsic mechanism terms $\beta_{1}^{\rm SHE}$ is much larger 
than $\beta_{1}^{\rm AHE}$. 
Asymmetric spin-dependent scatterings in the spin-flip processes induced by 
the electron-magnon interactions would be a possible explanation 
for the strong temperature dependence of the SHE in contrast to the AHE 
or even the SHE in nonmagnetic metals.

\acknowledgments 
We thank S. Maekawa and B. Gu for fruitful discussions. 
This work is supported by the Japanese Grant-in-Aid for Scientific Research 
on Innovative Area, ``Nano Spin Conversion Science" (Grant No. JP26103002), 
and by the Spanish MINECO under the Maria de Maeztu Units of 
Excellence Programme (MDM-2016-0618) and under 
Projects No. MAT2015-65159-R and MAT2017-82071-ERC. 
Y. Omori acknowledges financial support from Japan Society for the Promotion of Science (JSPS)
through ``Research program for Young Scientists" and 
``Program for Leading Graduate Schools (MERIT)". 
M.G. acknowledges financial support from the Leverhulme Trust 
via an Early Career Research Fellowship (ECF-2013-538). 
E.S thanks the Spanish Ministry of Education, Culture and Sport 
for a Ph.D. fellowship (Grant No. FPU14/03102).


\begin{thebibliography}{99}
\bibitem{ref}
The scattering probability related to the spin-flip process is defined 
as $P^{+-}_{kk'}\sim\left|\left<\Psi^{-}_{k'}\right|\hat{T}\left|\Psi^{+}_{k}\right>\right|^2$ 
where $\Psi^{+ (-)}_{k (k^{\prime})}$ is the wave function before (after) the scattering with the wave number 
$k (k^{\prime})$ and spin-up $+$ (spin-down $-$), and 
$\hat{T}$ is a matrix in spin space including spin-orbit coupling and magnetism. 
By defining $P^{-+}_{kk'}$ in an equivalent way and integrating $k$ 
and $k^\prime$, we find that on average it holds 
$\tau^{-1}_{+-}=\tau^{-1}_{-+}$, with $\tau^{-1}=\int \mathrm{dk}\int 
\mathrm{dk}^\prime P_{kk^\prime}$, even in the magnetic case.
\end{thebibliography}
\end{document}